# An ultra-fast superconducting Nb nanowire single-photon detector for soft X-rays

K. Inderbitzin,[1] A. Engel,[1] A. Schilling,[1] K. Il'in,[2] and M. Siegel[2]

[1] *Physics Institute, University of Zurich, Winterthurerstr. 190, 8057 Zurich, Switzerland*

[2] *Institute of Micro- and Nano-Electronic Systems, Karlsruhe Institute of Technology, Hertzstr. 16, 76187 Karlsruhe, Germany*

(Dated: 12 September 2012)

Although superconducting nanowire single-photon detectors (SNSPDs) are well studied regarding the detection of infrared/optical photons and keV-molecules, no studies on continuous X-ray photon counting by thick-film detectors have been reported so far. We fabricated a 100 nm thick niobium X-ray SNSPD (an X-SNSPD), and studied its detection capability of photons with keV-energies in continuous mode. The detector is capable to detect photons even at reduced bias currents of 0.4%, which is in sharp contrast to optical thin-film SNSPDs. No dark counts were recorded in extended measurement periods. Strikingly, the signal amplitude distribution depends significantly on the photon energy spectrum.

Already more than a decade before the development of superconducting nanowire single-photon detectors (SNSPD) for the optical and near-infrared wavelength range, serious efforts had been undertaken to adapt this detection principle for X-ray photons with keV-energies.[1,2,3,4] However, these preliminary X-ray detectors struggled with latching, making it difficult to operate them as self-recovering detectors in which superconductivity recovers after photon detection events (called continuous operation mode in this letter). The need to externally reduce the bias current to a value low enough for superconductivity to recover after a detection event results in long dead times and limits the count rates.

Very fast and sensitive X-ray single-photon detectors from superconducting nanowires would be very interesting for applications where very high count rates, precise timing, a good signal-to-noise ratio and response in a wide spectral range for photon counting are required. Potential applications comprise experiments with synchrotron X-ray sources, free-electron lasers and hot plasmas (as in nuclear fusion experiments), all emitting bright and pulsed X-ray radiation. In many medical imaging systems ultrafast X-ray single-photon detectors with energy resolution are desirable in order to reduce patient radiation dose. In the recently developing photon-counting X-ray computer tomography for example, long detection pulse durations can compromise the image quality by a possible overlap of succeeding photon pulses, as high photon fluxes have to be used in order to prevent motion blur.[5]

Recently, SNSPDs[6] and superconducting stripline detectors (SSLDs)[7,8,9,10] with superconducting film thicknesses of up to 50 nm were reported to be used in continuous mode for time-of-flight mass spectrometry (TOF-MS) of molecules with keV-energies, and ultrafast pulse recovery times and pulse rise times down to 380 ps ± 50 ps were reported.[11] There is only one report on continuous X-ray photon counting with SNSPDs: Perez de Lara *et al.*[12] reported on the detection of 6 keV photons by a SNSPD from 5 nm thin NbN. However, there are no published studies on continuous X-ray photon detection in thick-film SNSPDs (called X-SNSPDs in this work), although they are promising candidates for ultrafast detectors in continuous mode.

The absorptance of X-ray photons in thin-film superconductors is extremely low, in contrast to the underlying, commonly much thicker substrate. It was shown[12] that the 6 keV photon detection in thin-film SNSPDs mostly occurs through photon absorption in the substrate and the successive energy diffusion to the superconducting meander structure. In the reported case, this even leads to a device detection efficiency (DDE) which is higher than the photon absorption probability of the superconductor. The DDE is defined as the number of photon signal events, normalized to the number of photons crossing the active detector area. It is very likely, however, that this indirect detection principle for X-ray photons results in an increased timing jitter compared to optical photon detection, due to the random nature of the energy diffusion processes to the superconducting meander structure after photon absorption and the depth distribution of the absorption event. In order to enhance the absorptance of the superconducting detector and therefore promoting a more direct detection principle, we fabricated X-SNSPDs from a 100 nm thick niobium film. The absorptance of a 100 nm thick layer of niobium is 2.7% for 6 keV photons, and 0.23% for 30 keV photons.[13]

A niobium film of 100 nm thickness was grown at room temperature by DC magnetron sputtering in an Ar atmosphere on R-plane cut sapphire substrates. The critical temperature $T_c$ of the as-grown film was 8.9 K. From this film X-SNSPDs were fabricated using electron-beam lithography of a multi-layer ZEP520A resist and reactive ion etching, mostly adapting previously published fabrication recipes.[14] After fabrication, the detector showed a critical temperature $T_c = 8.4 \text{ K}$. Fig. 1(a) shows an optical image of the X-SNSPD with which the measurements in this report were performed. The meander covers an area of 131 x 55 μm$^2$ (meandering area) with a filling factor of about 50% by a conduction path of a uniform width $w \approx 250 \text{ nm}$ and a length $l \approx 11 \text{ mm}$. The accuracy in determining the width of the conduction path was limited by the optical resolution of the microscope. The superconducting leads to this meander structure have a length negligible for the total length $l$ and are much wider ($w_{Leads} \approx 4 \text{ μm}$) than the conduction path of the meander structure.

This detector geometry ensures that the detector has a kinetic inductance small enough to allow for ultrafast recovery times,[15] and is large enough to reduce problems with latching as previously reported,[1,3,4,16] therefore allowing its operation in continuous mode. In order to estimate the kinetic inductance $L_K$ of the X-SNSPD, a literature value[17] for the penetration depth of $\lambda(0) \approx 100 \text{ nm}$ is used for the 100 nm thick niobium film, and we assume $\lambda(1.75 \text{ K}) \approx \lambda(0)$, as the reduced temperature $t = T/T_c \approx 0.2$ is sufficiently small at the measurement temperature. In this case of a niobium film with a thickness $d = 100 \text{ nm} \approx \lambda$ the kinetic inductance can be approximated by $L_K = \mu_0 \lambda l / w$,[18] resulting in $L_K \approx 6 \text{ nH}$. This kinetic inductance is smaller than the values reported[9] for functional SSLDs from thinner films with 40 nm thick niobium, therefore, the tendency for latching of our X-SNSPD is expected to be larger. However, it will be shown, that the X-SNSPD can be operated in continuous mode.

The X-ray detection characterization measurements have been performed in a He-3 bath cryostat with a temperature stabilized at 1.75 K ± 0.05 K during the measurements. The detector signal was transmitted to a cryogenic amplifier at the 4 K-stage, and then further to a second amplifier at room temperature before fed into a 3.5 GHz digital oscilloscope. The amplifier chain had an effective bandwidth of about 40 MHz to 1.9 GHz. The bias current was applied in constant-voltage mode and was passed through a series of low-pass filters. The oscilloscope allowed for the recording of thousands of triggered voltage signals, hence a subsequent detailed signal analysis in order to distinguish between detector pulses (including possible dark counts) and unexpected noise signals with much higher voltage amplitude was performed for all measurements, which is not possible with the otherwise widely used threshold counters.

The X-SNSPD was irradiated by X-ray photons through a 100 μm thin window of polyimide (Kapton) at the cryostat, in order to minimize X-ray attenuation. A second window of 1 μm thin aluminum was installed at the 4K-heat shield to minimize thermal radiation. Two qualitatively different X-ray sources were used: A tungsten-target X-ray tube with a maximum acceleration voltage $V_A = 49.9\,\text{kV}$ and a radioactive Fe-55 isotope source with an activity of 3.7 GBq, which emits photons with well defined energies around 6 keV. The emission characteristics of the X-ray sources and the geometry of the experimental setup allowed a homogeneous irradiation of the X-SNSPD by both sources.

Fig.1(b) shows several examples of voltage pulses after X-ray photon absorption recorded by the oscilloscope, with the detector biased at $I_{\text{bias}} \approx 0.61\,\text{mA}$, which corresponds to about 5.5% of the experimentally measured critical current $I_c \approx 11\,\text{mA}$. The damped oscillations following the pulse are a consequence of the fact that the detector is effectively part of an LC circuit.[21] Defining the pulse length time $T_P$ as the period of the first pulse oscillation, i.e. the time span between the first and the third zero crossing, ultrafast recovery times of 3.4 ns ± 0.2 ns were measured under irradiation of the X-SNSPD with the X-ray tube at $V_A = 49.9\,\text{kV}$. The pulse amplitude $A$ is defined as the absolute value of the first pulse minimum (shown in Fig. 1(b)) and hence the rise time as the time span between the absolute values $10\% \cdot A$ and $90\% \cdot A$. We obtained a rise time of 250 ps ± 70 ps, which is lower than any reported value for TOF-MS with SSLDs.[11] Furthermore, we can expect the intrinsic signal rise time to be even faster, since the effective bandwidth $BW$ = 1.9 GHz of our electronics setup results in a theoretical minimum rise time $T_R \approx 0.35/BW \approx 190\,\text{ps}$.[19]

X-ray photon detection in this X-SNSPD was possible at bias currents as low as 0.4% $I_c$ [Fig. 1(c)]. This stands in sharp contrast to the behavior of SNSPDs for optical photons, even in the case of SNSPDs with especially low cut-off energy.[20,21] This can be explained by the fact that the detected photons have about $r_E \approx 10^3$ times higher energy than visible photons, and that at the same time the cross-section of the conduction path of this detector is only $r_A \approx 30$ times larger compared to a 7.5 nm thin niobium SNSPD in a reported work[22] on optical photon detection. A threshold value of $I_{bias}/I_c \approx 40\%$ can be extracted from their data, which we define as the reduced bias current for which the DDE is two orders of magnitude smaller than the maximum. For our X-SNSPD the threshold value for $I_{bias}/I_c$ can therefore be estimated to be $r_E/r_A \approx 30$ times smaller, which gives $I_{bias}/I_c \approx 1\%$. At $V_A = 30\,\text{kV}$ a threshold value of $I_{bias}/I_c \approx 2\%$ is extracted from our measurements. This simple estimate cannot be applied for the detection of X-ray photons in thin-film SNSPDs, where photon detection was only reported down to relatively high bias-currents of the order of 70% $I_c$,[12] as only a part of the photon energy is expected to diffuse into the superconducting meander after photon absorption in the substrate.

A second estimate for the threshold of $I_{bias}/I_c$ can be made with a hot-spot model following Gabutti et al.[1] for the detection of X-ray photons in superconducting strips from thick niobium films. They assume that the primary excitations of low keV-photons are created in a small volume centered at the absorption point of the photon, thereby creating a hot-spot resembling a sphere, in which superconductivity is destroyed. The bias current is therefore forced to flow in a smaller cross-section, where it can exceed $I_c$ for high enough bias currents, which leads to a normal conducting domain. For the absorption of a 8.4 keV photon (of the strong characteristic emission peak for $V_A = 30\,\text{kV}$, see below and Fig. 2), the hot-spot radius can be estimated as $r_0 \approx 210\,\text{nm}$ using Eq. (3) in Ref. 1, Eq. (2) in Ref. 4, the BCS expression for the heat capacity of low-temperature superconductors (Eq. (22) in Ref. 23) and the electronic specific heat constant for niobium.[24] Using Eq. (1) in Ref. 1 results in a vanishing reduced bias current threshold, as the hot-spot is expected to cover the whole cross-section. A more detailed look at the absorption process (see below) shows that this is a very crude approximation. On the other hand, the observed reduced bias current threshold results in a hot-spot radius of $r_0 \approx 120\,\text{nm}$. However, the actual reduced bias current threshold could be even lower as observed, as the count rate will also decrease for lower bias currents due to a decreasing signal amplitude while keeping the trigger level constant. Therefore this calculated hot-spot radius $r_0$ is only a lower limit, and the actual hot-spot could as well cover the whole cross-section.

Above a reduced bias current $I_{bias}/I_c \approx 5.5\%$, the device cannot be continuously used for X-ray photon detection, as the photon-induced normal conducting domain does not recover its superconducting state due to self-heating. This latching effect[16,25,26] is promoted by the low kinetic inductance $L_K$ of the X-SNSPD and the orders-of-magnitude smaller resistance of the normal conducting domains as compared to the situation in thin-film SNSPDs from NbN, originating from the larger conductance, cross-section and diffusivity of the niobium meander.

However, investigations on possible dark counts could be performed up to $I_{bias}/I_c = 99\%$, but no dark counts during more than 5 hours of measurement time were recorded, thereby limiting the dark-count rate to $\leq 5.4 \times 10^{-5}$ s$^{-1}$. The fact that the dark-count rate is much lower in our X-SNSPD compared to typical thin-film SNSPDs[27] is not surprising considering the geometric differences. Assuming thermally activated fluctuations to be responsible for dark counts, then the energy associated with phase-slips due to fluctuations in the superconducting order parameter scales at least with the cross-section of the superconductor $wd$, and the vortex-antivortex scenario is topologically suppressed for a film thickness $d >> \xi$. Also, the typical energy required for a single vortex to overcome the edge barrier is proportional to the film thickness,[28] thus reducing the probability to exceedingly small values for all three fluctuation events considered to contribute to the dark-count rate in thin-film SNSPD.

Fig. 1(d) shows the count rate dependence on the X-ray photon flux which was varied by the X-ray tube anode current $I_A$ at $V_A = 49.9$ kV, keeping the relative photon energy distribution constant.[29] This dependence shows a linear behavior, and we therefore conclude that the X-SNSPD detects in single-photon mode.[30]

The sample pulses in Fig. 1(b) show that the amplitudes $A$ of different pulses vary significantly, much more than in the case of SNSPDs.[31] In Fig. 2 we plotted histograms of the pulse amplitudes at $I_{bias} \approx 0.61$ mA for photons from the X-ray tube at different $V_A$ and from the radioactive Fe-55 source, shown only for amplitudes right above the noise level (amplitudes below the noise level could not be recorded). We attribute this amplitude variation to the small resistance of the normal conducting domains, which we estimate to be of the order of 1 Ω, using the estimated values for the hotspot radius from above as the half domain length. Since this value is smaller than the 50 Ω impedance of the coaxial signal line, the voltage drop over the detector varies with the resistance and thus the size of the normal conducting domains. In thin-film SNSPDs however, the domain resistances are of the order of a few hundred Ohms,[32] hence the voltage drop cannot vary to the same degree as in the examined detector.

Van Vechten *et al.*[33] examined in detail the processes which happen after absorption of a keV-photon in niobium: Initially, the X-ray photon energy is transferred to a primary photoelectron, which then scatters inelastically to produce secondary electrons causing the path of the primary to be irregularly kinked. The distance the primary electron will have travelled before becoming indistinguishable from the other electrons in the system (called primary range) depends on its initial energy, and is estimated to be about 170 nm for 6 keV and 2.8 μm for 30 keV in niobium. This process takes place within 1 ps or less, and another bound of $\approx 10^{-14}$ s can be calculated by assuming a uniform deceleration from the initial kinetic energy to zero velocity. These electrons create excited quasiparticles, which eventually relax to energies of a few meV and finally recombine into Cooper pairs within 1 ps to 1 ns if no bias current is applied, similarly to the quasiparticle diffusion in SNSPDs. This timescale is orders of magnitude longer than the time necessary for the primary to lose its energy. Hence, the emerging hot-spot (in which superconductivity is destroyed) will not have a spherical shape in most cases, as assumed by the hot-spot model by Gabutti *et al.*[1] discussed above, but rather the shape of a growing quasiparticle cloud along the random paths of the primary and secondary electrons.

Our examined meander is 100 nm thick and $\approx 250$ nm wide, which is of the same order of magnitude as the primary range of a 6 keV photoelectron given above. Assuming a photon is absorbed in the meander, different fractions of its energy will therefore be absorbed in the meander for different absorption events due to the random primary path (even for identical absorption points). This leads to different resistances of the emerging normal conducting domain and explains why the Fe-55 emitted photons lead to significantly varying pulse amplitudes [Fig. 2], despite their well-defined photon energy. At this point, indirect detection events originating from photons absorbed in the substrate close enough to the interface of the meander cannot be excluded. However, estimates of the absorption probability in 100 nm Nb and of the detection efficiency let us conclude that direct detection events dominate.

Nevertheless, we expect that two absorbed photons with different energies on average deposit different amounts of energy in the meander, therefore leading to different average sizes of normal conducting domains. Strikingly, Fig. 2 shows an amplitude distribution which depends significantly on $V_A$. Since the detector operates in a single-photon detection mode, this dependence can be attributed to the variation of the photon energy spectrum by $V_A$.

In order to demonstrate the correlation between the recorded pulse heights and the emission spectrum of the X-ray tube we have overlaid the spectrum (dashed red line, top and right axes) for $V_A \approx 30\,\text{kV}$,[34] linearly scaled to the amplitude distribution with a common origin. Additionally, we plotted the relative absorption probability of a photon with a certain energy taking into account the transmission through the two cryostat windows and the energy dependent absorption in 100 nm thick Nb (solid red line). For $V_A \geq 12.5\,\text{kV}$, our X-SNSPD shows distinct preferred signal amplitudes (see vertical red arrows in Fig. 2), which we may tentatively ascribe to the two main characteristic emission lines of the tungsten target at 8.3-8.4 keV and 9.7-10.0 keV.[35] These preferred signal amplitudes do not appear for $V_A \leq 10\,\text{kV}$, since $V_A \geq 10.2\,\text{kV}$ and $\geq 11.5\,\text{kV}$ are conditions for the excitation of the different emission lines. Additionally, the maximum amplitudes scale approximately linearly with $V_A$ up to about 12.5 kV. For higher voltages the amplitude distribution appears to be truncated, probably due to the mean propagation distance of the primary becoming much longer than the film thickness.

We note that a certain pulse amplitude variation has also been observed for superconducting stripline detectors from 40 nm thick niobium films detecting molecules with keV-energies,[9] where it was ascribed to supercurrent variations among parallel striplines and multiple impact events.

In conclusion, our results show that ultrafast dark-count-free X-SNSPDs from 100 nm thick niobium can be fabricated which can operate in a large spectral range. Using a hot-spot model, the X-ray photon detection capability at very low reduced bias currents is explained. The small resistance of the normal conducting domains leads to significant variations in the pulse amplitude, which are susceptible to the photon energy spectrum. This raises the question of a potential energy resolution of X-SNSPDs. The kinetic inductance of the device has to be carefully chosen by proper design, in order to reduce the tendency of the detector to latch due to the small domain resistances. An increase of the DDE of X-SNSPDs by additional fabrication steps (e.g. by stacking of several active superconducting layers as previously suggested for SNSPDs[36]) or by using other superconducting materials could be possible.

This research received support from the Swiss National Science Foundation Grant No. 200021_135504/1 and is supported in part by DFG Center for Functional Nanostructures under sub-project A4.3.

**REFERENCES**

[1]A. Gabutti, R. G. Wagner, K. E. Gray, R. T. Kampwirth, and R. H. Ono, Nucl. Instrum. Methods A **278**, 425 (1989).

[2]A. Gabutti, K. E. Gray, R. G. Wagner, and R. H. Ono, Nucl. Instrum. Methods A **289**, 274 (1990).

[3]A. Gabutti, K. E. Gray, G. M. Pugh, and R. Tiberio, Nucl. Instrum. Methods A **312**, 475 (1992).

[4]L. Parlato, G. Peluso, G. Pepe, R. Vaglio, C. Attanasio, A. Ruosi, S. Barbanera, M. Cirillo, and R. Leoni, Nucl. Instrum. Methods A **348**, 127 (1994).

[5]L. Yu, X. Liu, S. Leng, J. M. Kofler, J. C. Ramirez-Giraldo, M. Qu, J. Christner, J. G. Fletcher, and C. H. McCollough, Imaging Med. **1**, 65 (2009).

[6]K. Suzuki, S. Miki, Z. Wang, Y. Kobayashi, S. Shiki, and M. Ohkubo, J. Low Temp. Phys. **151**, 766 (2008).

[7]K. Suzuki, S. Miki, S. Shiki, Z. Wang, and M. Ohkubo, Appl. Phys. Express **1**, 031702 (2008).

[8]A. Casaburi, N. Zen, K. Suzuki, M. Ejrnaes, S. Pagano, R. Cristiano, and M. Ohkubo, Appl. Phys. Lett. **94**, 212502 (2009).

[9]N. Zen, A. Casaburi, S. Shiki, K. Suzuki, M. Ejrnaes, R. Cristiano, and M. Ohkubo, Appl. Phys. Lett. **95**, 172508 (2009).

[10]K. Suzuki, S. Shiki, M. Ukibe, M. Koike, S. Miki, Z. Wang, and M. Ohkubo, Appl. Phys. Expr. **4**, 083101 (2011).

[11]A. Casaburi, M. Ejrnaes, N. Zen, M. Ohkubo, S. Pagano, and R. Cristiano, Appl. Phys. Lett. **98**, 023702 (2011).

[12]D. Perez de Lara, M. Ejrnaes, A. Casaburi, M. Lisitskiy, R. Cristiano, S. Pagano, A. Gaggero, R. Leoni, G. Golt'sman, and B. Voronov, J. Low Temp. Phys. **151**, 771 (2008).

[13]B.L. Henke, E.M. Gullikson, and J.C. Davis, Atomic Data and Nuclear Data Tables Vol. **54** (no.2), 181-342 (1993).

[14]H. Bartolf, K. Inderbitzin, L. B. Gómez, A. Engel, and A. Schilling, J. Micromech. Microeng. **20**, 125015 (2010).

[15]A. J. Kerman, E. A. Dauler, W. E. Keichler, J. K. W. Yang, K. K. Berggren, G. Gol'tsman, and B. Voronov, Appl. Phys. Lett. **88**, 111116 (2006).

[16]B. V. Estey, J. A. Beall, G. C. Hilton, K. D. Irwin, D. R. Schmidt, J. N. Ullom, and R. E. Schwall, IEEE Trans. Appl. Supercond. **19**, 382 (2009).

[17]A. I. Gubin, K. S. Il'in, S. A. Vitusevich, M. Siegel, and N. Klein, Phys. Rev. B **72**, 064503 (2005).

[18]P. Müller and A. V. Ustinov, The Physics of Superconductors (Springer, Berlin, 1997), p. 39.

[19]N. S. Nise, *Control Systems Engineering* (Wiley, Hoboken NJ, 2004), p. 179-180.

[20]B. Baek, A. E. Lita, V. Verma, and S. W. Nam, Appl. Phys. Lett. **98**, 251105 (2011).

[21]A. Engel, A. Aeschbacher, K. Inderbitzin, A. Schilling, K. Il'in, M. Hofherr, M. Siegel, A. Semenov, and H.-W. Hübers, Appl. Phys. Lett. **100**, 062601 (2012).

[22]A. J. Annunziata, D. F. Santavicca, J. D. Chudow, L. Frunzio, M. J. Rooks, A. Frydman, and D. E. Prober, IEEE Trans. Appl. Supercond. **19**, 327 (2009).

[23]M. A. Biondi, A. T. Forester, M. P. Garfunkel, and C. B. Satterthwaite, Rev. Mod. Phys. **30**, 1109 (1958).

[24]D. R. Lide (Editor), *CRC Handbook of Chemistry and Physics* (CRC Press, Boca Raton FL, 2005), p. 2142.

[25]A. J. Kerman, J. K. W. Yang, R. J. Molnar, E. A. Dauler, and K. K. Berggren, Phys. Rev. B **79**, 100509(R) (2009).

[26]A. J. Annunziata, O. Quaranta, D. F. Santavicca, A. Casaburi, L. Frunzio, M. Ejrnaes, M. J. Rooks, R. Cristiano, S. Pagano, A. Frydman, and D. E. Prober, J. Appl. Phys. **108**, 084507 (2010).

[27]H. Bartolf, A. Engel, A. Schilling, K. Il'in, M. Siegel, H.-W. Hübers, and A. Semenov, Phys. Rev. B **81**, 024502 (2010).

[28]L. N. Bulaevskii, M. J. Graf, C. D. Batista, and V. G. Kogan, Phys. Rev. B **83**, 144526 (2011).

[29]Oxford Instruments data sheet for the X-ray tube used in the experiments described in this letter, *XTF5011 Flux Data* (received 4/2010).

[30]G. N. Gol'tsman, O. Okunev, G. Chulkova, A. Lipatov, A. Semenov, K. Smirnov, B. Voronov, A. Dzardanov, C. Williams and R. Sobolewski, Appl. Phys. Lett. **79**, 705 (2001).

[31]A. Semenov, P. Haas, K. Il'in, H.-W. Hübers, M. Siegel, A. Engel, and A. Smirnov, Physica C **460**, 1491 (2007).

[32]A. D. Semenov, P. Haas, B. Günther, H.-W. Hübers, K. Il'in, M. Siegel, A. Kirste, J. Beyer, D. Drung, T. Schurig, and A. Smirnov, Supercond. Sci. Technol. **20**, 919 (2007).

[33]D. Van Vechten and K. S. Wood, Phys. Rev. B **43**, 12852 (1990).

[34]Oxford Instruments Application Note: *Typical Spectrum For Tungsten-Target X-ray tube* (AN012 Rev. A, revised 3/2007).

[35]A. Thompson, D. Attwood, E. Gullikson, M. Howells, K.-J. Kim, J. Kirz, J. Kortright, I. Lindau, Y. Liu, P. Pianetta, A. Robinson, J. Scofield, J. Underwood, G. Williams, and H. Winick, *X-Ray Data Booklet* (Lawrence Berkeley National Laboratory, Berkeley, 2009), p. 1-12 and p. 1-21.

[36]A. Matthew Smith, in *Proceedings of SPIE Volume 8400*, Baltimore, Maryland United States, 26-27 April 2012, edited by E. Donkor, A. R. Pirich, and H. E. Brandt (SPIE, Bellingham, 2012), 84000L.

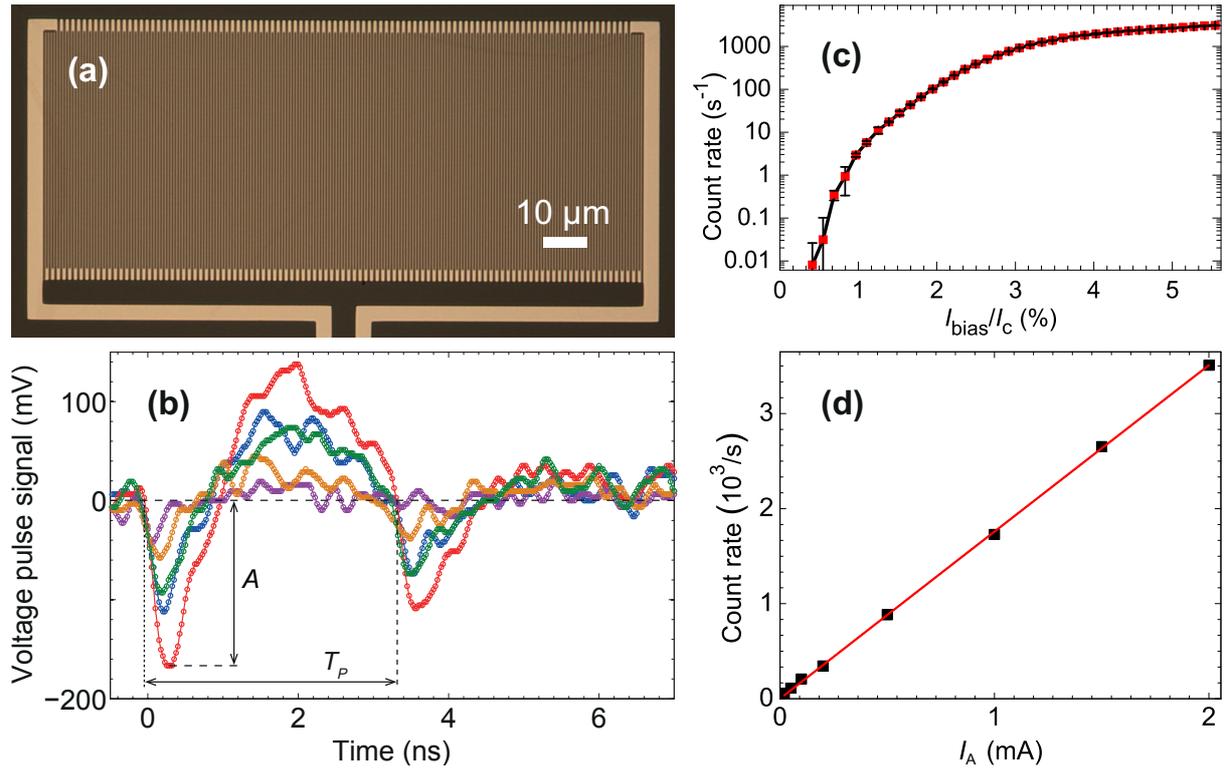

FIG. 1. (a) Optical image of examined X-SNSPD from 100 nm thick niobium. For (b)-(d) the X-SNSPD was irradiated by the X-ray tube at $V_A = 49.9\,\text{kV}$, which emits at maximum intensity at $I_A = 2.00\,\text{mA}$. (b) Typical voltage pulses after photon absorption, with the definition of the pulse length $T_P$ and the pulse amplitude $A$ shown schematically. The different colors of the pulses are only a guide to the eye. (c) Photon count rate as a function of $I_{\text{bias}}/I_c$. (d) Photon count rate as a function of $I_A$, which is proportional to the photon flux, showing a linear dependence and thus single-photon detection.

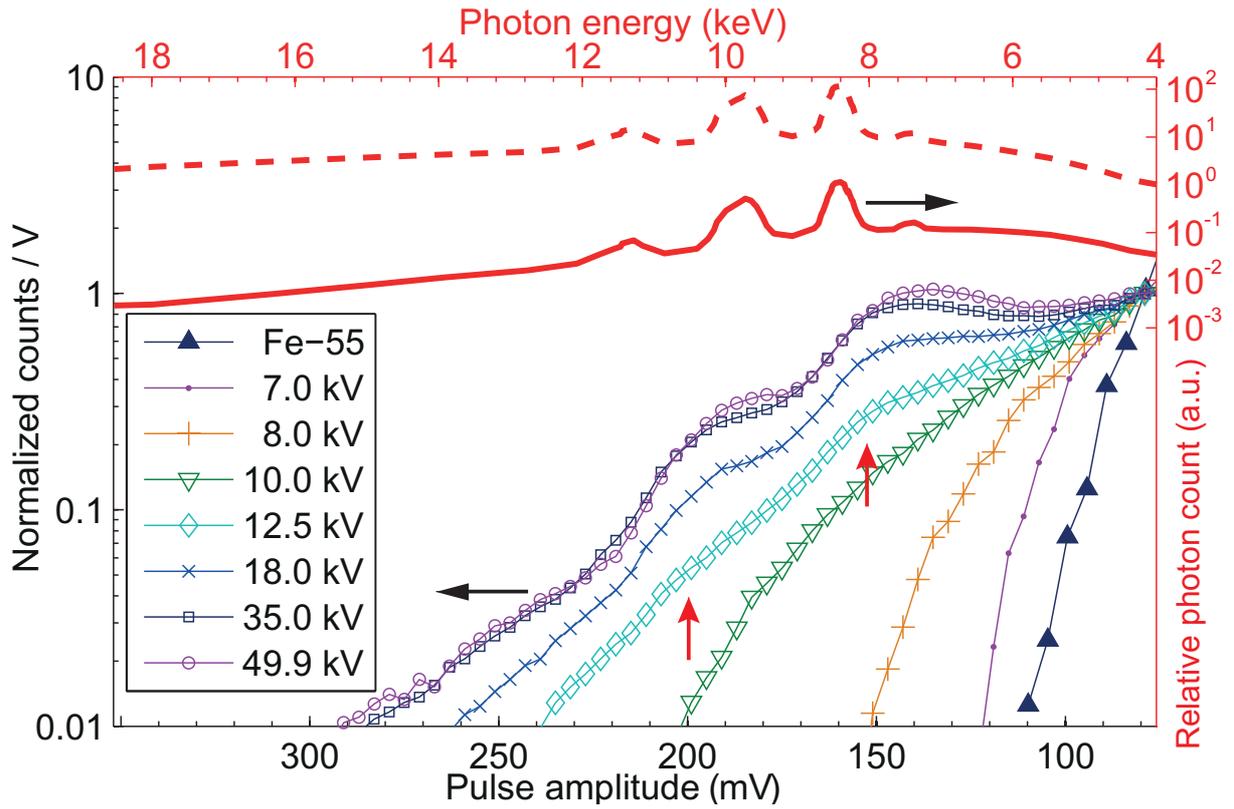

FIG. 2. The top and right axes refer to the X-ray tube spectrum at $V_A \approx 30\,\text{kV}$ (dashed red line). Taking into account the transmission through the cryostat windows and the absorption in 100 nm niobium, the solid red line shows the relative probability for the absorption of a photon. The left and bottom axes refer to histograms of voltage pulse amplitudes from photons emitted by the X-ray tube at different $V_A$ (indicated in the legend, $V_A$ determines the maximum energy of the tube emitted photons) and by the radioactive Fe-55 source, which mainly emits at 6 keV. The histograms use a bin size of 4 mV (5.2 mV for the Fe-55 data) and are normalized at 79 mV, which lies above the noise level. The two vertical red arrows indicate preferred signal amplitudes which may tentatively be ascribed to the main characteristic emission at 8.3-8.4 kV and 9.7-10.0 kV.